\documentclass[10pt,showpacs,amsmath,amssymb,floatfix,superscriptaddress,longbibliography,twocolumn,eqsecnum]{revtex4-2}
\usepackage{amsmath}
\usepackage{physics}
\usepackage{graphicx}
\usepackage[usenames,dvipsnames,svgnames,table]{xcolor}
\usepackage{tcolorbox}
\usepackage{booktabs}
\usepackage{makecell} 
\usepackage{tabularx}
\usepackage{chngcntr}
\usepackage[utf8]{inputenc}
\counterwithout{equation}{section}
\counterwithout{figure}{section}
\usepackage[unicode=true,
            pdfusetitle,
            bookmarks=true,
            bookmarksnumbered=false,
            bookmarksopen=false,
            breaklinks=true,
            pdfborder={0 0 0},
            backref=false,
            colorlinks=true,
            hypertexnames=false]{hyperref}
\hypersetup{linkcolor=NavyBlue,urlcolor=NavyBlue,citecolor=NavyBlue}

\begin{document}
	
\title{Nonlinear dynamics of intra-cavity-field for periodically pump-modulated dual-coupled Kerr micro-ring cavities}
	
\author{Yiyi Zhang}
\affiliation{College of Physics, Hangzhou Dianzi University, Hangzhou 310018, China}
\author{Chaoying Zhao}
\email{zchy49@163.com}
\affiliation{College of Physics, Hangzhou Dianzi University, Hangzhou 310018, China}
\affiliation{State Key Laboratory of Quantum Optics Technologies and Devices, Shanxi University, Taiyuan, 030006,China}
\affiliation{ Zhejiang Key Laboratory of Quantum State Control and Optical Field Manipulation, Hangzhou Dianzi University, Hangzhou, 310018, China}
\date{\today}

\begin{abstract}
We numerically investigate the long-time intra-cavity-field dynamics of a periodically pump-modulated pair of coherently coupled Kerr micro-ring resonators. Along a detuning scanning, fixed-phase stroboscopic trajectories change from recurrent regular state to states with positive maximum-Lyapunov growth. Long observation windows near the transition, where finite-time estimates converge slowly. The low- and high-detuning states both retain multi-mode Kerr-comb spectra, while their cycle-to-cycle field organization changes much more strongly. Spatiotemporal intensities, adjacent-cycle profiles, and full-field recurrence measures show a loss of one-period recurrence together with changes in the inter-resonator phase relation, coherent exchange, and dispersive intensity flow. We linearize the coupled equations around the long-time field and examine the growth of an infinitesimal perturbation. The resulting tangent-energy balance shows that the loss contribution changes little along the scanning, whereas the Kerr contribution increases and exceeds the loss magnitude in the high-detuning states with positive Lyapunov growth. A separate dimensional correspondence relates the normalized working range to $Si_3N_4$ micro-ring. These results provide a field-resolved description of the long-time dynamical change and its associated perturbation growth.
\end{abstract}

\maketitle

\section{Introduction}
Kerr micro-ring resonators provide a compact platform for driven-dissipative nonlinear optics and optical frequency-comb generation. In a coherently pumped cavity, loss is compensated by an external field, detuning controls the phase relation between the pump and the resonant mode, group-velocity dispersion introduces mode-dependent phase accumulation, and the Kerr effect produces an intensity-dependent nonlinear phase shift. These ingredients are naturally combined in the Lugiato-Lefever equation (LLE), which connects the slow evolution of the intra-cavity envelope with its fast-time structure and discrete modal spectrum~\cite{Lugiato1987,Chembo2013,Hansson2013,Godey2014,Coen2013}. The framework supports modulational instability, localized structures, multi-stability, and complex long-time states~\cite{Herr2014,Zhang2025}.

With two coupled Kerr resonators, the fields can also exchange energy and phase between the cavities. Previous work on dual-coupled optical cavities has shown that inter-resonator coupling, detuning, and free-spectral-range mismatch can strongly modify continuous-wave states, modulational instability, and comb formation~\cite{Zhu2022}. More generally, coupled nonlinear cavities and photonic dimers support self-pulsing, chaos, symmetry breaking, and mode hybridization through the combined action of dissipation, Kerr nonlinearity, and coherent coupling~\cite{Maes2009,Grigoriev2011,Yelo2021,Yelo2022,Ghosh2023,Woodley2021,Sanyal2025}. Dual-ring devices have been used experimentally to control the effective coupling and comb-generation efficiency, while strongly coupled Kerr dimers have exhibited a broad range of nonlinear states and mode dynamics~\cite{Miller2015,Tikan2021,Tikan2022}. Recent work has examined soliton interaction, bound-state formation, and perturbation sensitivity in coupled Kerr micro-resonators~\cite{Dolinina2026}.

Pulsed and amplitude-modulated driving has been used to control localized Kerr-cavity states, repetition rates, and synchronization dynamics~\cite{Hendry2018,Lobanov2015,Lobanov2019,Mizrahi2024}, while a periodic drive provides an explicit external time scale in a dissipative Kerr cavity~\cite{Taheri2022}. Recent theory of di-chromatically driven Kerr resonators has analyzed multi-stability, synchronization, and chaos in dissipative cavity-soliton dynamics~\cite{Taheri2025}. We consider a weak periodic modulation of the pump intensity. The period defines a natural stroboscopic section on which we compare the fields $F_j(nT_m,\vartheta)$ and $F_j[(n+1)T_m,\vartheta]$ at the same modulation phase. Successive field snapshot can be compared without mixing the internal dynamics with a change of the external pump phase.

 We compare the comb spectrum and the cycle-to-cycle intra-cavity fields when the two detunings are scanned together while their difference and the coupling strength are fixed. The local intensity balance is used to separate phase, coherent-exchange, and dispersive-flow contributions. Due to finite-time Lyapunov estimates fluctuate near the transition, a longer post-transient calculation is used to distinguish the two representative endpoints from the intermediate states. The normalized parameter is related to a representative $Si_3N_4$ device scale. Finally, we linearize the coupled equations around the long-time field and compare the loss and Kerr contributions to the perturbation growth.

\section{Theoretical Model}
\begin{figure}[t]
\centering
\includegraphics[width=\columnwidth]{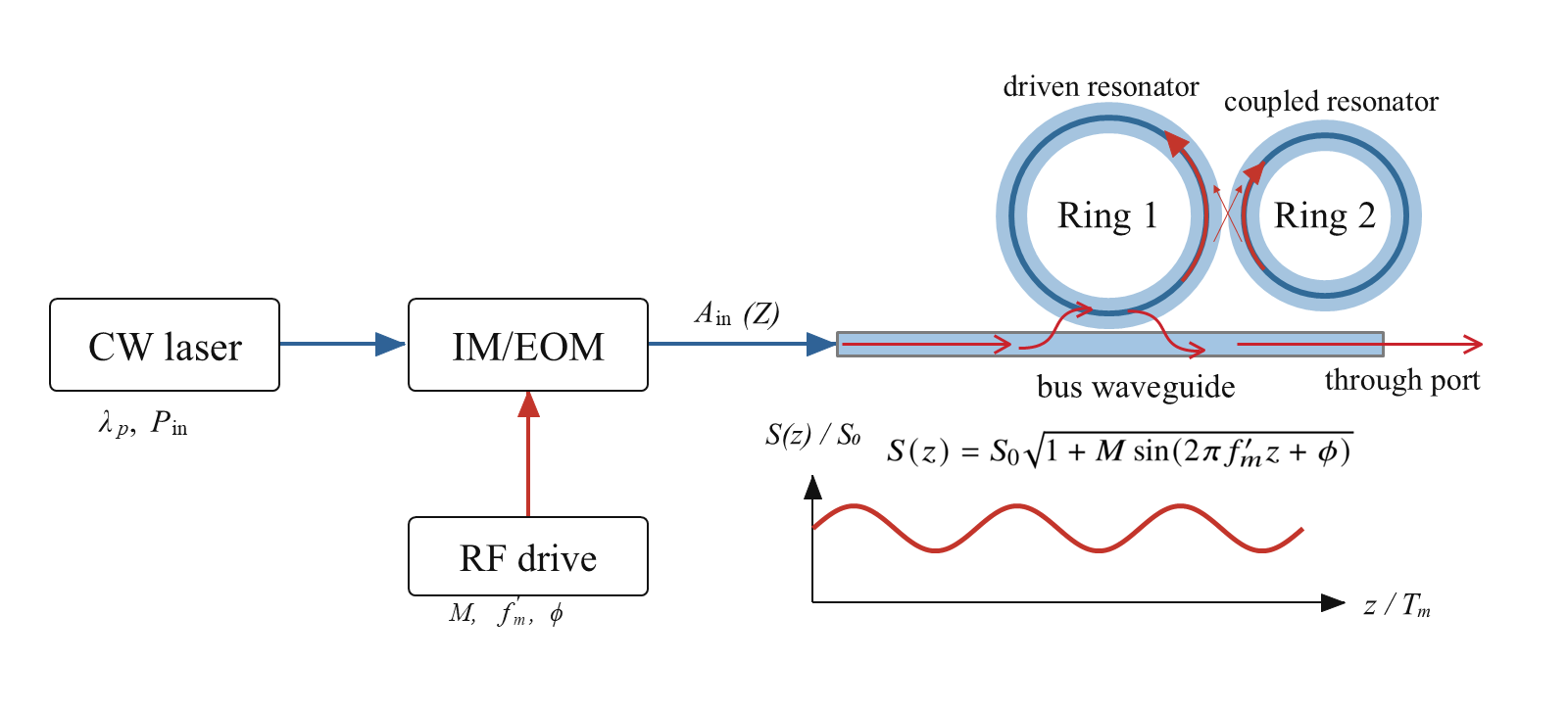}
\caption{Schematic of the periodically pump-modulated dual-coupled Kerr micro-ring cavity. A continuous-wave(CW) laser is intensity modulated by an radio frequency (RF)-driven intensity modulator(IM)/electro-optic modulators (EOM) before driving Ring~1, while Ring~2 is excited through the inter-ring coupler. The normalized pump field injected into Ring~1 is $S(z)=S_0\sqrt{1+M\sin(2\pi f'_m z+\phi)}$.}
\label{fig:system}
\end{figure}

We consider two Kerr micro-ring coupled coherently to each other. Ring~1 is coupled to a bus waveguide and is driven by an external coherent field. Ring~2 receives energy and phase information only through the inter- field coupling. Figure~\ref{fig:system} illustrates the system and the periodic modulation. Let $Z$ denote the slow evolution coordinate and $\tau$ is the fast time in the co-moving frame. The slowly varying envelopes in the two resonators are denoted by $A_1(Z,\tau)$ and $A_2(Z,\tau)$. The dimensional mean-field equations are
\begin{align}
\frac{\partial A_1}{\partial Z}={}&(-\alpha_1-i\delta_1-\Delta k\frac{\partial}{\partial\tau}-i\frac{\beta_{21}}{2}\frac{\partial^2}{\partial\tau^2}+i\gamma_1|A_1|^2)A_1 \nonumber\\
&+i\kappa_c A_2+i\kappa_1 A_{in}(Z),\label{eq:dim1en}\\
\frac{\partial A_2}{\partial Z}={}&(-\alpha_2-i\delta_2-i\frac{\beta_{22}}{2}\frac{\partial^2}{\partial\tau^2}+i\gamma_2|A_2|^2)A_2+i\kappa_c A_1.\label{eq:dim2en}
\end{align}
Here $\alpha_j$ are the field-loss rates, $\delta_j$ are the pump-cavity detunings, $\beta_{2j}$ are the second-order dispersion coefficients, $\gamma_j$ are the Kerr coefficients, $\kappa_c$ is the coherent inter-resonator coupling rate, $\kappa_1$ is the external field-coupling coefficient between the bus waveguide and Ring~1, and $\Delta k$ accounts for the free-spectral-range mismatch. This coupled Lugiato-Lefever equation (LLE) structure follows the standard mean-field description of dual-coupled Kerr micro-ring cavities~\cite{Zhu2022}.

We use
\begin{equation}
z=\alpha_2 Z, \vartheta=\tau\sqrt{\frac{2\alpha_2}{|\beta_{22}|}}, F_j=A_j\sqrt{\frac{\gamma_2}{\alpha_2}},
\end{equation}
and define
\begin{gather}
\alpha=\frac{\alpha_1}{\alpha_2}, \Delta_j=\frac{\delta_j}{\alpha_2},
 d=\Delta k\sqrt{\frac{2}{|\beta_{22}|\alpha_2}},\\
\eta_1=\frac{\beta_{21}}{|\beta_{22}|}, \eta_2=\frac{\beta_{22}}{|\beta_{22}|},
\Gamma=\frac{\gamma_1}{\gamma_2}, \kappa=\frac{\kappa_c}{\alpha_2}.
\end{gather}
The normalized periodic pumping is
\begin{equation}
S(z)=S_0\sqrt{1+M\sin(2\pi f'_m z+\phi)},T_m=1/f'_m.
\end{equation}
The normalized coupled equations are 
\begin{equation}
\begin{aligned}
\frac{\partial F_1}{\partial z}={}&(-\alpha-i\Delta_1-d\partial_\vartheta-i\eta_1\partial_\vartheta^2+i\Gamma|F_1|^2)F_1\\
&+i\kappa F_2+i S(z),
\end{aligned}
\label{eq:lle1en}
\end{equation}
\begin{equation}
\frac{\partial F_2}{\partial z}=(-1-i\Delta_2-i\eta_2\partial_\vartheta^2+i|F_2|^2)F_2+i\kappa F_1.
\label{eq:lle2en}
\end{equation}
We use
\begin{equation}
\alpha=1.1, d=0,\eta_1=\eta_2=-1, \Gamma=1,\kappa=1.2,
\end{equation}
\begin{equation}
S_0=2.8, M=0.02, f'_m=0.25, T_m=4.
\end{equation}
The representative values are $u=0.100,0.125,0.150,0.175,0.200$ with $u=0.160,0.165, 0.170$ used for the transition analysis, respectively.
The detunings follow
\begin{equation}
\Delta_1=2.4+u, \Delta_2=2.2+u,
\end{equation}
thus the pump-resonance offset is varied while $\Delta_1-\Delta_2=0.2$ remains fixed.

Table~\ref{tab:mapping} gives out loaded quality factors, dispersion, and Kerr coefficients for free spectral range (FSR) Silicon Nitride( $Si_3N_4$) dual-ring with $\lambda_0=1555.37nm$, group index $n_g=2.11$, circumference $L=473.606~\mu m$, and micro-ring radius $R=75.377~\mu m$.

\begin{table}[t]
\caption{Normalized simulation parameters and a representative dimensional correspondence for a Free Spectral Range (FSR) Silicon Nitride( $Si_3N_4$) dual-ring.}
\label{tab:mapping}
\centering
\scriptsize
\setlength{\tabcolsep}{2.5pt}
\begin{tabular}{lll}
\toprule
Quantity & Value & Representative scale\\
\midrule
$\alpha$ & 1.1 & $Q_{L,1}\simeq1.55\times10^7$, $Q_{L,2}\simeq1.70\times10^7$\\
$d$ & 0 & $\mathrm{FSR}_{1,2}\simeq300GHz$\\
$\eta_1,\eta_2$ & $-1,-1$ & $\beta_{2,1,2}\simeq-141.1ps^2km$\\
$\Gamma$ & 1 & $\gamma_1\simeq\gamma_2\simeq 0.7W$ $^{-1}$ $m^{-1}$\\
$\kappa$ & 1.2 & cavity splitting $\simeq 13.61MHz$\\
$S_0$ & 2.8 & pumping $\simeq0.365mW$\\
$M$ & 0.02 & $2\%$ intensity modulation\\
$f'_m$ & 0.25 & $f_m\simeq8.905MHz$\\
$u$ & $0.10-0.20$ & detuning shift $\simeq0.567MHz$\\
\bottomrule
\end{tabular}
\end{table}

We define the ordinary-frequency detuning scale
\begin{equation}
\nu_0=\frac{\alpha_2 L\mathrm{FSR}}{2\pi}\simeq5.669~\mathrm{MHz}.
\label{eq:dimscale}
\end{equation}
The dimensional conversions in Table~\ref{tab:mapping} follow from the definitions of the normalized quantities,
\begin{equation}
\Delta\nu_j=\Delta_j\nu_0,
f_m=2\pi f'_m\nu_0,
\Delta\nu_{split}=2\kappa\nu_0.
\label{eq:dimconvert}
\end{equation}
The factor $2\pi$ in the modulation-frequency conversion comes from the phase $2\pi f'_m z$ in the periodic drive, whereas the factor of two in $\Delta\nu_{split}$ denotes the full separation of the two hybridized cavity super-modes.  $f'_m=0.25$ gives out $f_m\simeq8.905MHz$ and $\kappa=1.2$ gives out a splitting of approximate to $13.61MHz$, while the scanning $u=0.10-0.20$ spans approximate to $0.567MHz$. The periodic driving can be produced by the RF-driven IM in Fig.~\ref{fig:system}, while the detuning path corresponds to a laser-frequency sweep over the coupled resonances. The dimensional values in Table~\ref{tab:mapping} are design targets derived from the same normalization.

To separating the roles of the Kerr, dispersive, and coupling terms, we written out
\begin{equation}
F_j(z,\vartheta)=\sqrt{I_j(z,\vartheta)}e^{i\phi_j(z,\vartheta)},I_j=|F_j|^2.
\end{equation}
The Kerr term is in phase quadrature with the local field, so the contribution to local intensity derivative vanishes,
\begin{equation}
\left.\partial_z I_j\right|_{Kerr}=0,
\end{equation}
whereas the contribution to the local phase evolution is
\begin{equation}
\left.\partial_z\phi_1\right|_{Kerr}=\Gamma I_1,
\left.\partial_z\phi_2\right|_{Kerr}=I_2.
\end{equation}
Thus, the Kerr nonlinearity produces an intensity-dependent local nonlinear phase shift.

With the Fourier convention
\begin{equation}
F_j(\vartheta)=\sum_\mu a_{j,\mu}e^{i\mu\vartheta}, \partial_\vartheta\rightarrow i\mu,
\end{equation}
and $\Delta\phi_{12}=\phi_2-\phi_1$, we define the coherent inter-resonator exchange term, the dispersion-induced local intensity flux, and the pumping contribution as
\begin{align}
J_{12}&=2\kappa\,\mathrm{Im}(F_1^*F_2)=2\kappa\sqrt{I_1I_2}\sin\Delta\phi_{12},\label{eq:j12en}\\
J_{{disp},j}&=-2\eta_j\,\mathrm{Im}(F_j^*\partial_\vartheta F_j)=-2\eta_j I_j\partial_\vartheta\phi_j,\label{eq:jdispen}\\
P_{pump}&=2\,\mathrm{Im}[S^*(z)F_1].\label{eq:pumpen}
\end{align}
With $\Delta\phi_{12}=\phi_2-\phi_1$, the sign convention in Eq.~\eqref{eq:j12en} gives out $J_{12}>0$ for net coherent intensity transfer from Ring~1 to Ring~2. This direction is consistent with the paired $-J_{12}$ and $J_{12}$ terms.
The local intensity equations become
\begin{align}
\partial_z I_1+\partial_\vartheta(dI_1+J_{{disp},1})&=-2\alpha I_1-J_{12}+P_{pump},\label{eq:balance1en}\\
\partial_z I_2+\partial_\vartheta J_{{disp},2}&=-2I_2+J_{12}.\label{eq:balance2en}
\end{align}
Eqs.~\eqref{eq:balance1en} and \eqref{eq:balance2en} separate the local terms used in the field-resolved analysis below. Kerr nonlinearity modifies the local phase; dispersion redistributes intensity through a fast-time flux; and coherent coupling transfers intensity between the two resonators through paired source and sink terms. When $d=0$, the drift term in Ring~1 vanishes.

The fast-time coordinate is represented spectrally. Linear loss, detuning, dispersion, and the $2\times2$ coupling block are applied in Fourier space, while the Kerr and pumping terms are updated in the physical domain, the related split-step and spectral schemes for nonlinear wave propagation~\cite{Sinkin2003}. The calculations use $N_\vartheta=128$ points and a slow-evolution step $dz=0.005$. After a $20$-period transient, each trajectory is followed for $1000$ modulation periods and per period at the same pumping phase. The full complex fields $F_1(nT_m,\vartheta)$ and $F_2(nT_m,\vartheta)$ are stored for the field-resolved analysis.

The maximum Lyapunov exponent is estimated with a Benettin perturbation in the full dual micro-ring complex-field space. The relative perturbation norm is $10^{-8}$ and is re-normalized per modulation period. The $1000$-period record is used to reduce the influence of short finite-time fluctuations near the transition. Representative calculations with a smaller slow-evolution step and a higher spectral resolution give out the same qualitative distinction between the regular endpoint and the high-detuning positive-Lyapunov state.

Due to the pumping modulation makes the equations non-autonomous in $z$, the relevant long-time object is the modulation-period map. We  characterize the dynamics on a stroboscopic section at a fixed modulation phase and use the maximum Lyapunov exponent of the full complex field as the primary diagnostic of long-time instability.

\section{Long-Time Dynamics and Kerr-Comb States}
 At $u=0.150$, $\lambda_{\max}\simeq1.3\times10^{-5}~z^{-1}$ approaches to zero on the long observation window, so this state is used as the regular reference. The intermediate points $u=0.160$, $0.165$, and $0.175$ show stronger finite-time variations and are retained as transition states. By contrast, the estimates at $u=0.170$ and $0.200$ remain positive over the long record. The interval $u\simeq0.160-0.175$ is treated as a transition neighborhood.

\begin{figure}[t]
\centering
\includegraphics[width=\columnwidth]{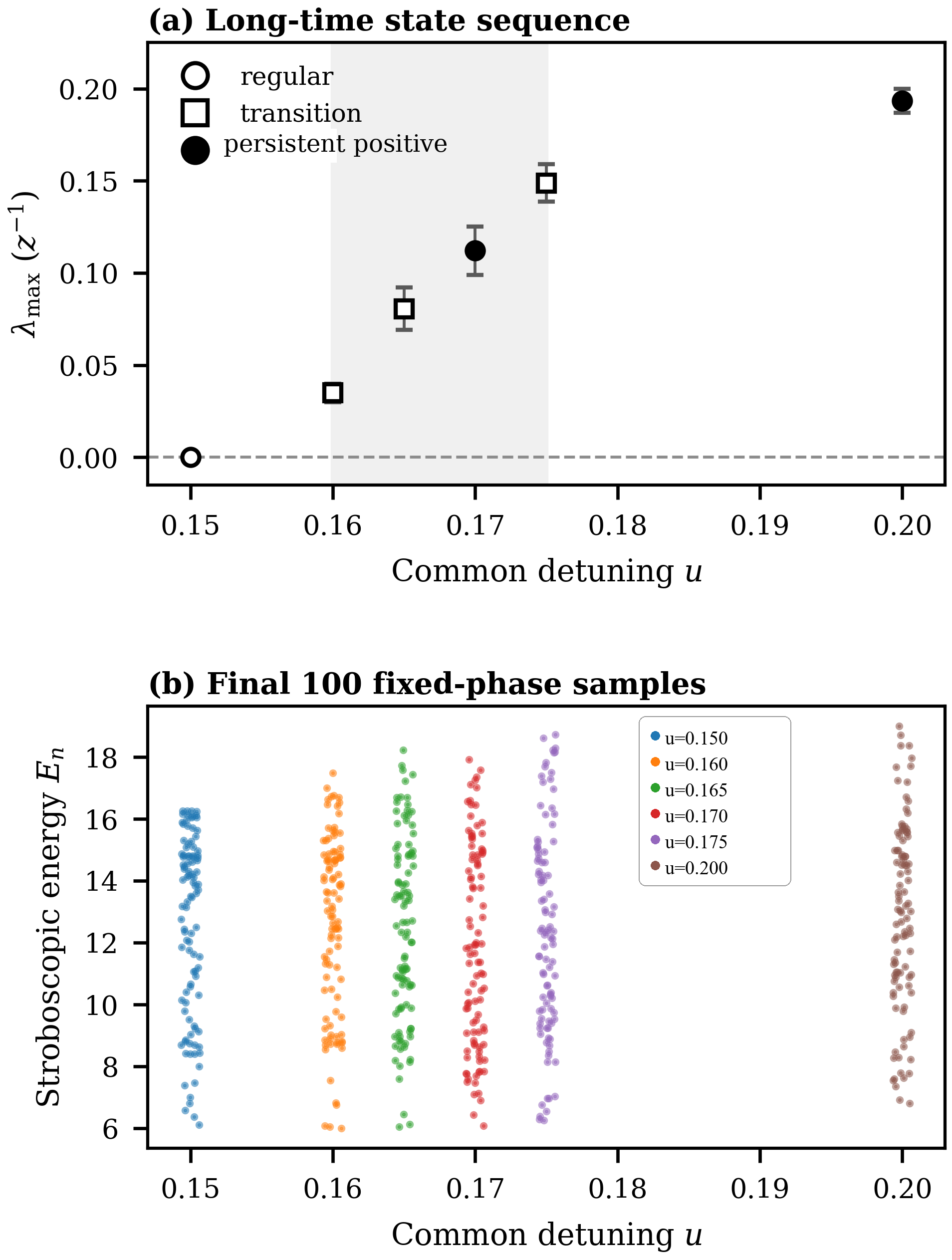}
\caption{Long-time state evolution along the common-detuning path. (a) Maximum-Lyapunov estimate from the $1000$-period post-transient record. Open circles, open squares, and filled circles denote regular, transition, and positive-Lyapunov states, respectively; the gray band marks the transition neighborhood. (b) Final $100$ stroboscopic Ring-1 energy shown for visual clarity; all $1000$ post-transient periods are used for the long-time statistics.}
\label{fig:state}
\end{figure}

\begin{table}[t]
\caption{Long-time states for $N_\vartheta=128$, $dz=0.005$, and $1000$ post-transient modulation periods. The listed $\lambda_{\max}$ values are obtained from the long records, the intermediate points as transition states due to their finite-time estimates vary more strongly along the observation window.}
\label{tab:longtime}
\centering
\scriptsize
\setlength{\tabcolsep}{20pt}
\begin{tabular}{rrl}
\toprule
$u$ & $\lambda_{\max}$ & State\\
\midrule
0.150 & 0.000013 & regular\\
0.160 & 0.034921 & transition\\
0.165 & 0.080581 & transition\\
0.170 & 0.112118 & positive-Lyapunov\\
0.175 & 0.148842 & transition\\
0.200 & 0.193395 & positive-Lyapunov\\
\bottomrule
\end{tabular}
\end{table}
The two endpoint states: $\lambda_{\max}$ approaches to zero at $u=0.150$, remains positive at $u=0.200$, where it is approximate to $0.1934~z^{-1}$. The stroboscopic energy strongly modulated even at low detuning, its spread is not used by itself to classify the long-time state. It instead records how the cycle-resolved energy organization changes as shown in Fig.~\ref{fig:state}.

\begin{figure}[t]
\centering
\includegraphics[width=\columnwidth]{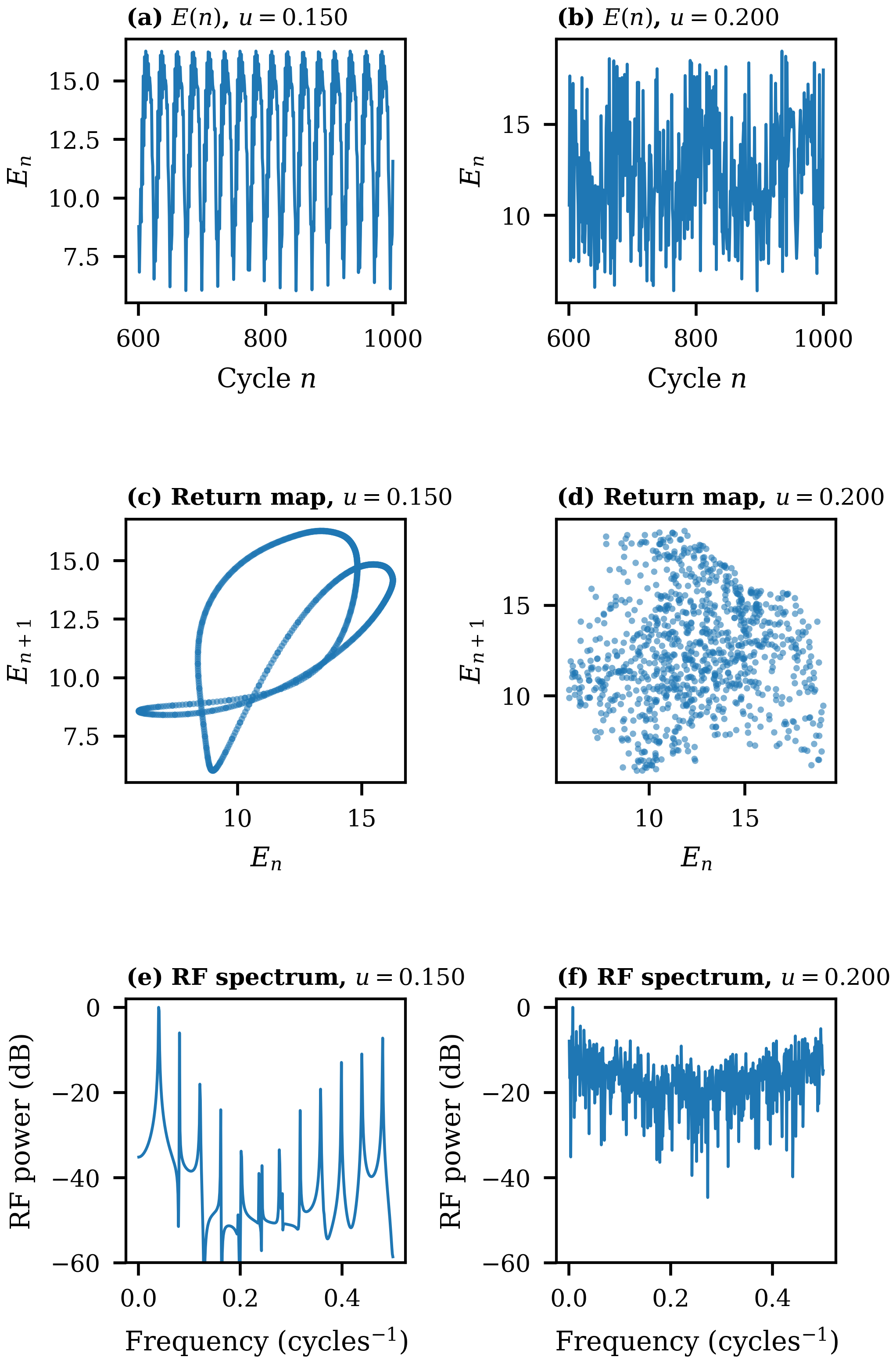}
\caption{Representative fixed-phase stroboscopic dynamics. (a) $E(n)$ for $u=0.150$; (b) $E(n)$ for $u=0.200$; (c,d) one-step return maps; and (e,f) RF spectra. The left and right columns correspond to the regular and high-detuning positive-Lyapunov  states, respectively.}
\label{fig:strobe}
\end{figure}

The regular state at $u=0.150$ and the high-detuning positive-Lyapunov state at $u=0.200$ provide two long-time anchors for the field comparison. At $u=0.150$, the energy sequence remains organized, the one-step return map is concentrated on a structured set, and the RF power is dominated by distinct components. At $u=0.200$, the return points spread over a broader two-dimensional region and the RF response becomes broadband as shown in Fig.~\ref{fig:strobe}. The contrast is obtained at the same modulation phase and reflects the long-time state.

\begin{figure}[t]
\centering
\includegraphics[width=\columnwidth]{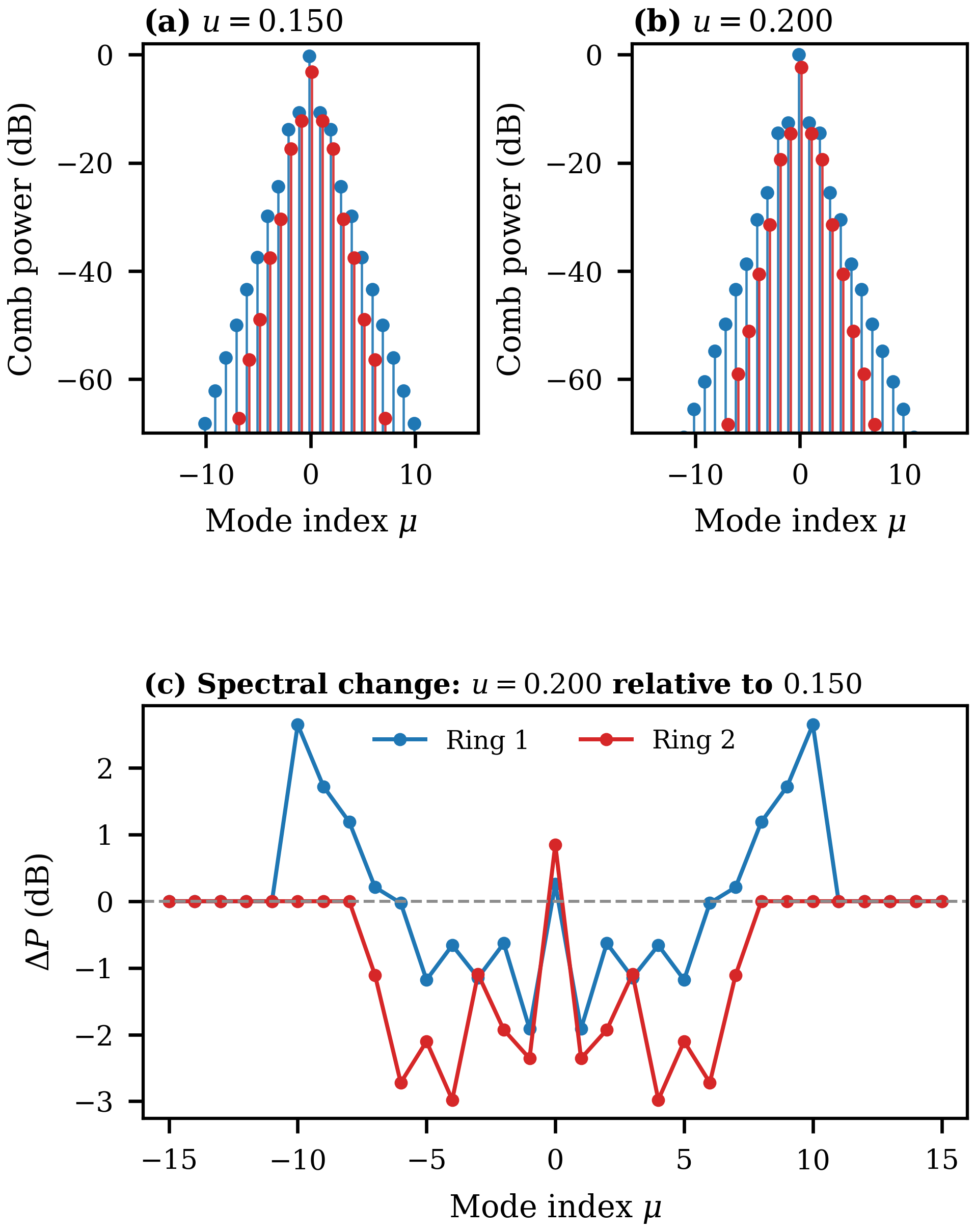}
\caption{Cycle-averaged optical Kerr-comb spectra. (a) $u=0.150$ and (b) $u=0.200$, with a normalization for resonators and states. The vertical lines emphasize the discrete comb teeth. (c) Mode-resolved spectral change $\Delta P=10\log_{10}[P_j(0.200)/P_j(0.150)]$, evaluated with the same $-70dB$ numerical floor used for the displayed comb window.}
\label{fig:comb}
\end{figure}

To verifying the modal content of the intra-cavity fields, the Fourier transform for the complex field at the fixed modulation phase,
\begin{equation}
a_{j,\mu}(n)=\frac{1}{2\pi}\int_{-\pi}^{\pi}F_j(nT_m,\vartheta)e^{-i\mu\vartheta}\,d\vartheta,
\end{equation}
and calculate the cycle-averaged modal power $\overline{P}_{j,\mu}=\langle|a_{j,\mu}(n)|^2\rangle_n$. The states at $u=0.150$ and $u=0.200$ both retain the pumping mode and multiple discrete sidebands as shown in Fig.~\ref{fig:comb}(a,b). Thus both  states are supported by multi-mode Kerr-comb fields.

The averaged comb envelope changes less strongly than the cycle-to-cycle dynamics. To making the spectral redistribution, Fig.~\ref{fig:comb}(c) plots the mode-resolved ratio $\Delta P=10\log_{10}[P_j(0.200)/P_j(0.150)]$. The stronger distinction between the two states remains in the temporal organization of the intra-cavity field.

\section{Local Intra-cavity-Field Dynamics}
\begin{figure}[t]
\centering
\includegraphics[width=\columnwidth]{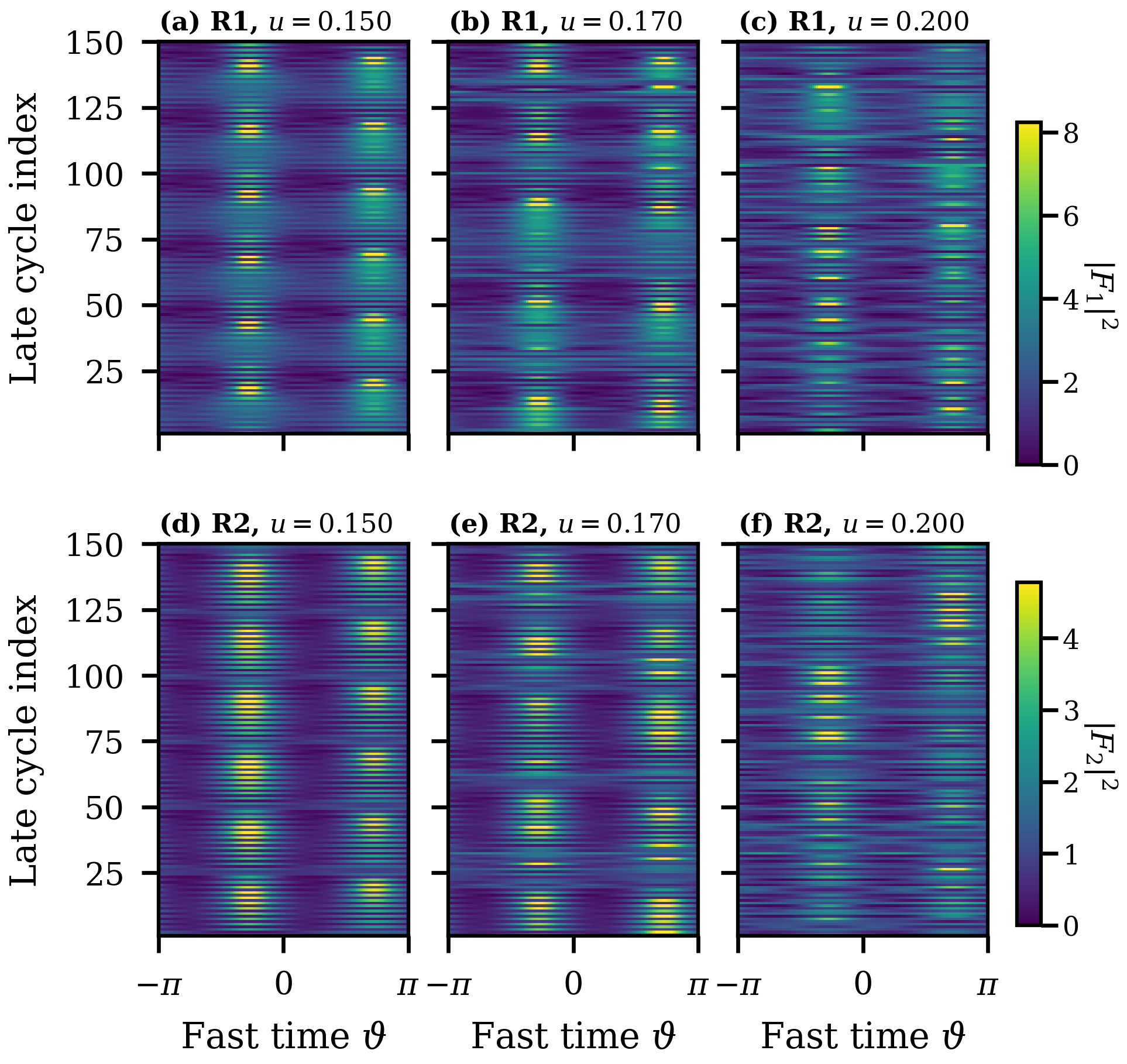}
\caption{Stroboscopic spatiotemporal intra-cavity-intensity distributions across the dynamical transition. The upper and lower rows correspond to Rings~1 and 2, respectively, while the three columns show $u=0.150$, $0.170$, and $0.200$. The horizontal axis is the fast-time coordinate $\vartheta$ and the vertical axis is the modulation-cycle index at a fixed modulation phase. The final $150$ post-transient cycles are displayed. The point $u=0.170$ is included as an intermediate state within the transition region. A color range is used across the three detunings within each resonator.}
\label{fig:intensity}
\end{figure}

Figure~\ref{fig:intensity} shows the stroboscopically intra-cavity intensities $I_j(nT_m,\vartheta)=|F_j|^2$ at three detunings. At $u=0.150$, both resonators retain an organized cycle-fast-time pattern, with the main high-intensity structures recurring at fast-time positions. At $u=0.170$, the cycle-to-cycle recurrence is reduced and both the local high-intensity regions and the background texture undergo stronger rearrangement. This point is included as an intermediate state within the transition region. At $u=0.200$, the positions, widths, and relative amplitudes of local structures vary much more strongly from one modulation period to the next.

\begin{figure}[t]
\centering
\includegraphics[width=\columnwidth]{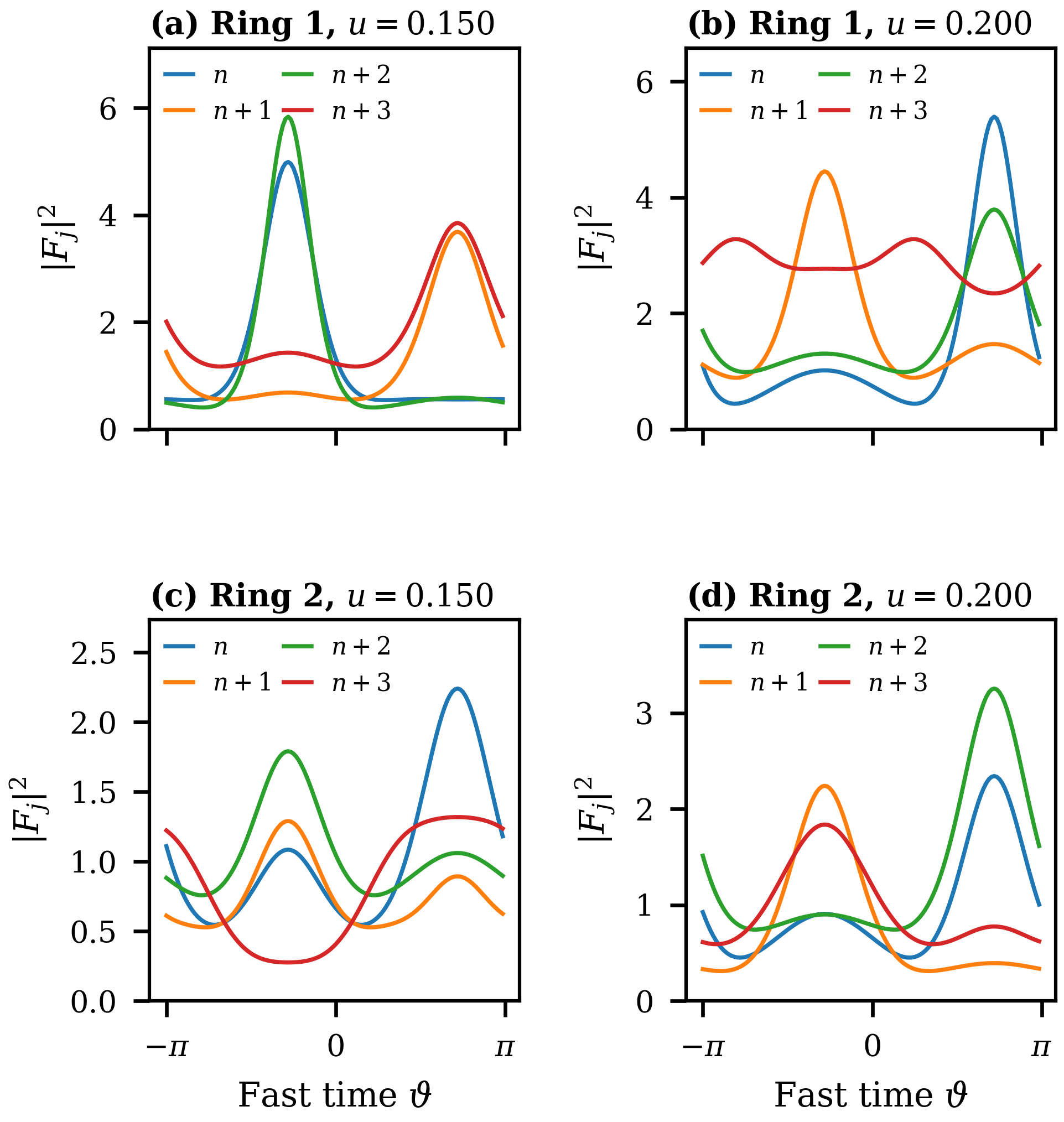}
\caption{Cycle-resolved intra-cavity intensity profiles on the two sides of the transition region. Panels (a,c) correspond to $u=0.150$, and panels (b,d) to $u=0.200$. Four consecutive modulation periods are overlaid in each panel to compare the recurrence of the field shape in the same resonator.}
\label{fig:profiles}
\end{figure}

\begin{figure}[t]
\centering
\includegraphics[width=\columnwidth]{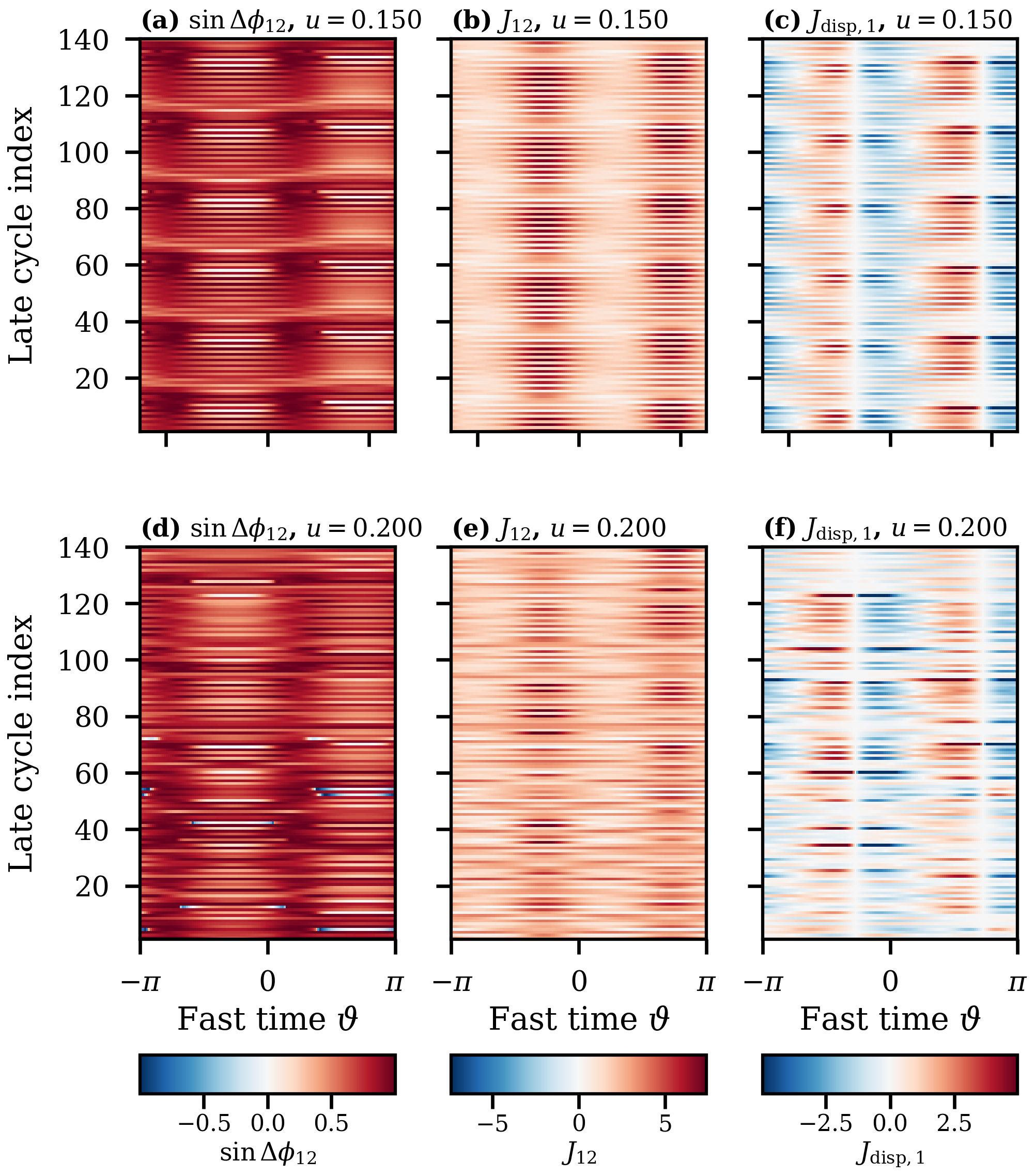}
\caption{Local phase, coherent inter-resonator exchange, and dispersion-induced intensity flow. The three columns show $\sin\Delta\phi_{12}$, $J_{12}=2\kappa\mathrm{Im}(F_1^*F_2)$, and $J_{{disp},1}=-2\eta_1\mathrm{Im}(F_1^*\partial_\vartheta F_1)$, respectively. The upper and lower rows correspond to $u=0.150$ and $u=0.200$, using the final $140$ post-transient cycles. The two states within each column share the zero-centered color range.}
\label{fig:local}
\end{figure}

The comparison maps the long-time change in Figs.~\ref{fig:state} and \ref{fig:strobe} onto the intra-cavity fields. Because the two resonators have different average intensity scales, the three detunings share the same color range. Figure~\ref{fig:profiles} resolves the same change through adjacent cycle-resolved fast-time profiles, while Fig.~\ref{fig:local} separates the local phase, coherent-exchange, and dispersive-flow contributions.

Figure~\ref{fig:profiles} overlays four consecutive stroboscopic intensity profiles to resolving the local-field changes on the two sides of the transition region. At $u=0.150$, the profiles in both resonators retain a clear shape. The dominant peaks remain at fast-time positions and their widths vary moderately; most of the cycle-to-cycle difference appears as a small modulation of the amplitudes. Because the four curves with the same pumping-modulation phase, this correspondence reflects a high recurrence of the intra-cavity intensity after one complete modulation period.

At $u=0.200$, the four profiles differ much more strongly. In Ring~1, the dominant peak and the background components are reorganized from cycle to cycle, while Ring~2 exhibits stronger amplitude and shape variations. These profiles show that the high-detuning state has a much lower one-period recurrence of the local intensity field. The displacement and deformation of individual peaks are local manifestations of this full-field reorganization.

The local intensity balance in Eqs.~\eqref{eq:balance1en} and \eqref{eq:balance2en} separates the coupling and dispersive contributions to the intensity evolution. The first column of Fig.~\ref{fig:local} shows $\sin\Delta\phi_{12}$, which contains the phase-difference factor controlling the sign of the coherent exchange. The second column shows $J_{12}$, whose magnitude depends on $\sqrt{I_1I_2}$. The phase factor and the coherent exchange carry related but distinct information. At low detuning, both quantities retain a comparatively organized pattern in the cycle direction, whereas at $u=0.200$ their local structures are rearranged much more strongly from period to period.

The third column of Fig.~\ref{fig:local} shows $J_{{disp},1}$. Since $J_{{disp},1}=-2\eta_1 I_1\partial_\vartheta\phi_1$, the fast-time phase gradient sets the local direction of the dispersive flow, while $I_1$ provides the corresponding intensity weight; the divergence $\partial_\vartheta J_{{disp},1}$ enters into the local intensity balance. At $u=0.200$, both $J_{12}$ and $J_{{disp},1}$ exhibit stronger cycle-to-cycle rearrangement. The same long-time change is accompanied by a reorganization of the local phase relation, coherent exchange, and dispersive transport of the intra-cavity field.

\section{Full-Field Recurrence and Perturbation Growth}
 We quantify above behavior with the normalized full-field distance
\begin{equation}
D_j(n)=\frac{\|F_j[(n+1)T_m]-F_j[nT_m]\|_2}{\|F_j[nT_m]\|_2},\label{eq:Den}
\end{equation}
and the normalized complex-field overlap
\begin{equation}
C_j(n)=\frac{|\langle F_j[nT_m],F_j[(n+1)T_m]\rangle|}{\|F_j[nT_m]\|_2\|F_j[(n+1)T_m]\|_2}.\label{eq:Cen}
\end{equation}
The two quantities describe complementary aspects of one-period recurrence. $D_j$ measures the complex-field displacement and is sensitive to a uniform phase shift, whereas the absolute overlap $C_j$ removes that global-phase sensitivity and emphasizes the normalized field shape and internal phase organization. Neither quantity is used to define chaos.

\begin{figure}[t]
\centering
\includegraphics[width=\columnwidth]{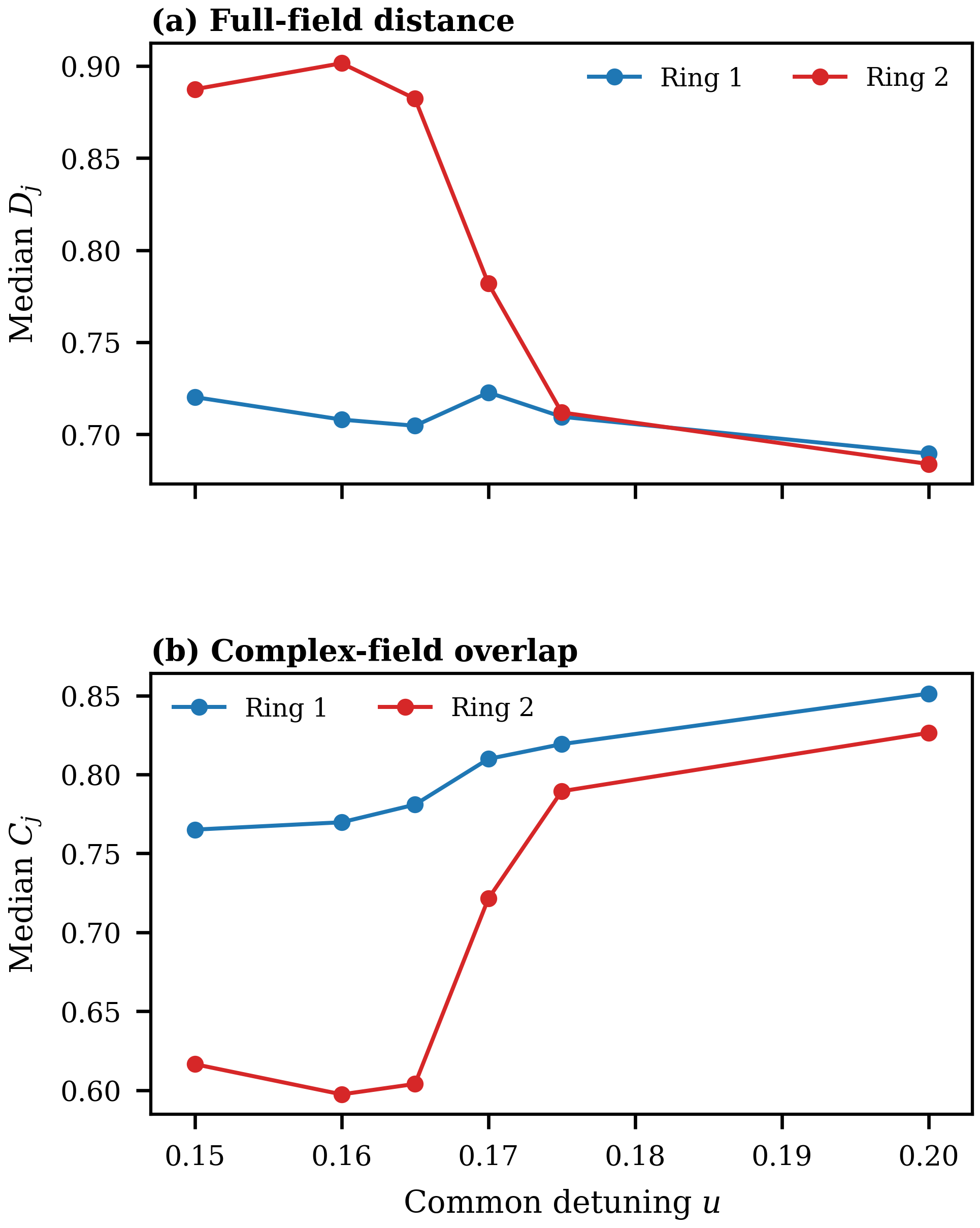}
\caption{Cycle-to-cycle recurrence of the full intra-cavity complex fields. (a) Median normalized field distance $D_j$ and (b) median normalized overlap $C_j$ for Rings~1 and 2. Connecting lines are guides to the eye between the sampled detunings.}
\label{fig:recurrence}
\end{figure}

The median $D_j$ and $C_j$ values vary non-monotonically in the transition neighborhood, but both rings show a clear rearrangement of one-period recurrence over the scanning as shown in Fig.~\ref{fig:recurrence}. Ring~2 is less recurrent at the low-detuning end and approaches to the Ring 1 values as $u$ increases. These measures provide the quantitative closure of the local-field analysis.

The field observables above establish how the transition appears in the long-time trajectory. We linearize the coupled LLE around the long-time field, $p_j=\delta F_j$. The tangent equations are
\begin{align}
\partial_z p_1={}&(-\alpha-i\Delta_1-d\partial_\vartheta-i\eta_1\partial_\vartheta^2
+2i\Gamma|F_1|^2)p_1\nonumber\\
&+i\Gamma F_1^2p_1^*+i\kappa p_2,\\
\partial_z p_2={}&(-1-i\Delta_2-i\eta_2\partial_\vartheta^2
+2i|F_2|^2)p_2\nonumber\\
&+i F_2^2p_2^*+i\kappa p_1.
\end{align}
The additive pumping does not enter into the tangent equations directly, it changes the long-time field $F_j$ and the coefficients of the linearized system.

\begin{figure}[t]
\centering
\includegraphics[width=\columnwidth]{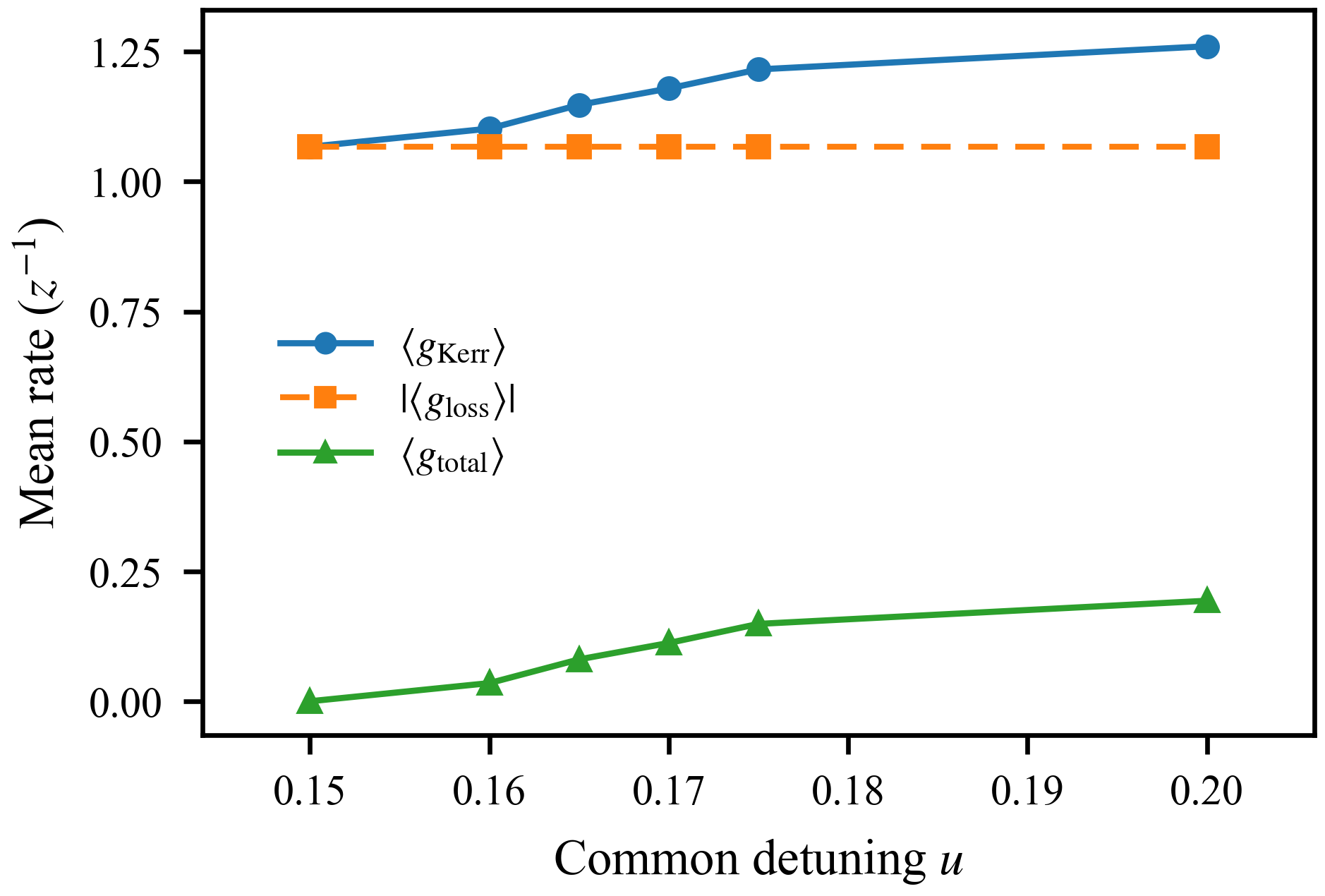}
\caption{Tangent-energy balance along the detuning scanning. The period-averaged Kerr tangent contribution, loss magnitude, and net tangent growth are shown for the states.}
\label{fig:tangent}
\end{figure}

To retaining the local interpretation used for the base field, define $q_j=|p_j|^2$ together with
\begin{align}
J_{\delta,12}&=2\kappa\,\mathrm{Im}(p_1^*p_2),\\
J_{\delta,{disp},j}&=-2\eta_j\,\mathrm{Im}(p_j^*\partial_\vartheta p_j),
\end{align}
and the Kerr tangent sources
\begin{equation}
G_{K,1}=2\Gamma\operatorname{Re}[i F_1^2(p_1^*)^2],
G_{K,2}=2\operatorname{Re}[i F_2^2(p_2^*)^2].
\label{eq:kerr_tangent_source}
\end{equation}
The local tangent-intensity balances are then
\begin{align}
\partial_z q_1+\partial_\vartheta(dq_1+J_{\delta,{ disp},1})&=-2\alpha q_1-J_{\delta,12}+G_{K,1},\\
\partial_z q_2+\partial_\vartheta J_{\delta,{ disp},2}&=-2q_2+J_{\delta,12}+G_{K,2}.
\end{align}
After integration over the periodic fast-time coordinate, the drift and dispersive fluxes vanish and the coherent-exchange terms cancel between the two micro-rings. Detuning and the diagonal Kerr phase terms preserve the total perturbation norm.
\begin{equation}
N_\delta=\int_{-\pi}^{\pi}(|p_1|^2+|p_2|^2)d\vartheta,
\end{equation}
the remaining terms give
\begin{equation}
g_{total}=\frac{1}{2}\frac{d\ln N_\delta}{dz}=g_{loss}+g_{Kerr},
\end{equation}
where
\begin{equation}
g_{loss}=-\frac{\alpha\|p_1\|^2+\|p_2\|^2}{N_\delta}.
\end{equation}
and
\begin{equation}
g_{Kerr}=\frac{\Gamma\operatorname{Re}[\!\int i F_1^2(p_1^*)^2d\vartheta]+\operatorname{Re}[\!\int i F_2^2(p_2^*)^2d\vartheta]}{N_\delta}.
\end{equation}
The period-averaged budget agrees with the independently calculated Benettin exponent over the states, the mismatch remains below $2.5\times10^{-4}~z^{-1}$. The loss contribution changes little with $u$, while the mean Kerr contribution increases and exceeds the loss magnitude at the high-detuning positive-Lyapunov states as shown in Fig.~\ref{fig:tangent}.

\section{Conclusion}
We have examined how a periodically driven dual-coupled Kerr micro-ring cavity changes along a detuning path from a recurrent regular state to long-time positive-Lyapunov dynamics. The multi-mode Kerr-comb structure persists on both sides of the transition, whereas the cycle-to-cycle organization of the intra-cavity field changes much more strongly. Energy and RF observables, spatiotemporal intensity maps, adjacent-cycle profiles, and the full-field measures $D_j$ and $C_j$ consistently trace this loss of one-period recurrence. The local intensity balance shows that the same change is accompanied by a reorganization of the inter-resonator phase relation, coherent exchange, and dispersive intensity flow.

The tangent-field calculation connects these field-level signatures to the growth of nearby trajectories. Dissipative loss remains nearly unchanged along the scanning, while the Kerr tangent contribution increases until the mean perturbation growth becomes positive at the high-detuning positive-Lyapunov states. Within the detuning path, the results connect the long-time dynamical change with two features: a loss of cycle-to-cycle intra-cavity-field recurrence and a Kerr contribution to perturbation growth that becomes larger than the dissipative loss. 

\begin{acknowledgments}
		This work was funded by the State Key Laboratory of Quantum Optics Technologies and Devices, Shanxi University, Shanxi, China (Grants No.KF202503); Zhejiang Key Laboratory of Quantum State Control and Optical Field Manipulation, Hangzhou Dianzi University (KYZ074326001).
	\end{acknowledgments}
    
\section*{DATA AVAILABILITY}
The data that support the findings of this article are not publicly available. The data are available from the authors upon reasonable request.

\bibliography{ref}
\end{document}